\documentclass[preprint,fleqn,showpacs,showkeys]{revtex4}
\usepackage{graphicx}
\usepackage{amssymb}
\usepackage{amsmath}
\usepackage{bm}
\usepackage{soul,xcolor}
\begin{document}

\setstcolor{red}
\newcommand{\ad}[1]{{\sf\color[rgb]{1,0,0}{#1}}} 
\newcommand{\zj}[1]{{\sf\color[rgb]{0,0,1}{#1}}} 
\newcommand{\zd}[1]{{\color{red} \st{#1}}}
\newcommand{\zy}[1]{{\color{green} \st{#1}}}
\setstcolor{green}
\newcommand{\gr}[1]{{\sf\color[rgb]{0,1,0}{#1}}} 
\newcommand{\gst}[1]{{\color{green} \st{#1}}}
\newcommand{\bea}{\begin{eqnarray}}
\newcommand{\eea}{\end{eqnarray}}
\newcommand{\wt}{\widetilde}
\setcounter{page}{0}
\title[]{An exact solution 
of the higher-order gravity
\\in standard 
radiation-dominated era
}

\author{Chae-min Yun}
\email{clair.yun@gmail.com}

\author{Jubin \surname{Park}}
\email{honolov@ssu.ac.kr} 

\author{Myung-Ki Cheoun}
\email{cheoun@ssu.ac.kr}

\affiliation{
Department of Physics and Origin of Matter and Evolution of Galaxies (OMEG) Institute, 
Soongsil University, Seoul 06978, Republic of Korea}

\author{Dukjae Jang}
\email{djjang2@ibs.re.kr}
\affiliation
{Center for Relativistic Laser Science, Institute for Basic Science, Gwangju 61005, Republic of Korea}
\begin{abstract}
We 
report
that the standard evolution of radiation-dominated era (RDE) universe 
$a \propto t^{1/2}$ 
is a sufficient condition for solving a sixth order gravitational field equation derived from the Lagrangian containing $B R^{ab}R_{ab} + C R {R^{;c}}_{c}$ 
  as well as a polynomial $f(R)$
 for a spatially flat radiation 
FLRW universe. 
By virtue of the 
similarity between $R^{ab}R_{ab}$ and $R^2$ models up to the background order and of the vanishing property of ${R^{;c}}_{c}$ for $ H = 1/(2t)$, the analytical solution 
can be obtained
from a special case to general one. 
This proves that  
the standard cosmic evolution 
is valid
even within modified gravitational theory involving higher-order terms.
An application of this background solution to the tensor-type perturbation reduces the complicated equation to the standard second order equation of gravitational wave. 
%
We discuss the possible ways to discriminate the modified gravity model on the observations such as the gravitational wave from the disturbed universe and primordial abundances.

\end{abstract}



\maketitle

\section{INTRODUCTION}
Even though Einstein’s general relativity (GR) has successfully passed the observational tests in the solar system scale,
lots of efforts to generalize GR for cosmology
have been continued. One of them
is to introduce additional higher-order derivative terms due to both theoretical and phenomenological reasons. 
For instance, sixth order $R 
\Box R$ \cite{90 Gottloeber etal} as well as fourth order $R^{ab} R_{ab}$ terms \cite{97 N and H, 99 N and H, 10 Sotiriou and Faraoni, 82 birrell and davis, 05Mukhanov, Hawking and Ellis, 87 Starobinsky and Schmidt} are considered here with the action    
\bea
&& S    =  \int d^4x \sqrt{-g} \Big[ {1 \over 16\pi } \big( f(R) + B R^{ab} R_{ab} + C R  \Box R  \big)  + L_m \Big]  \label{action} ,  
\eea
where 
$f(R)$ is a polynomial 
 function of the Ricci scalar $R$ 
\bea
&& f(R) = \sum_{n=1}^{N} A_{(n)} R^n = R + A_{(2)} R^2 + A_{(3)} R^3 + ... , \quad A_{(1)} \equiv 1, \quad A_{(2)} \equiv A ,  \label{polynomial}
\eea 
$A_{(n)}, B, C$ are  constants,
$R_{ab}$ is the Ricci tensor. 
The d'Alembertian of $R$ is
$ 
\Box R \equiv g^{ab} R_{,a;b} = { R^{;c} }_{c}$
where the semicolon 
denotes covariant derivative,
and
the matter part Lagrangian is 
defined as
$\delta \big(\sqrt{-g}L_m  \big)  \equiv {1 \over 2} \sqrt{-g} T^{ab}_{{(m)}} \delta g_{ab}$.
%
In this paper
we follow the Hawking-Ellis \cite{Hawking and Ellis} convention, 
but 
adopt (Planckian) natural units
 $ c \equiv 1 \equiv G_{\mathcal{N}} \equiv \hbar \equiv k_B$. 
%
As special cases of general $f(R)$ gravity \cite{10 Sotiriou and Faraoni, 11 Nojiri and Odintsov,10 De Felice and Tsujikawa},
$N=1$ and $N=2$ in Eq.\,(\ref{polynomial}) correspond to the gravity theory of Einstein and Starobinsky \cite{80 Starobinsky}, respectively. The $R^2$ theory is not only favored by Planck Collaboration \cite{Planck2018} as an inflationary model that predicts successfully some observables 
such as the spectral index and tensor-to-scalar ratio, 
but also 
by scalaron dark matter model that estimates its mass \cite{21 Shtanov}. 
Moreover, there have been 
attempts to extend the 
$R^2$ theory.
The model with the
next $N=3$ term
as a small contribution to $N=2$ theory was also studied \cite{90 Berkin and Maeda, 20 Cheong etal, 22 rds etal}, recently, 
$N=4$ case was also investigated \cite{21 Aziz etal}.

Further, a fourth order $B R^{ab} R_{ab}$ theory, that is neither of $f(R)$ model nor conformally equivalent to Einstein gravity, 
was also introduced in the literature 
and textbooks
\cite{97 N and H, 99 N and H, 10 Sotiriou and Faraoni, 82 birrell and davis, 05Mukhanov, Hawking and Ellis, 87 Starobinsky and Schmidt}.
%
%
For the physical meaning or motivation of $CR \Box R$ theory, e.g., a conformal equivalence to two interacting scalar fields causing inflation,  
we refer to Ref.\,\cite{90 Gottloeber etal}.
According to a review for higher order gravity theory
\cite{07 Schmidt},
$\Box R$ model was pioneered by Buchdahl \cite{51 Buchdahl} (1951),
and quantum gravitational higher order corrections to Einstein-Hilbert action
was the idea of Sakharov \cite{67 Sakharov} (1967), prior to Starobinsky.
The case including $N=3$ and $CR \Box R$ without the $B$-term for an inflationary regime was studied using phase diagrams and conformal transformation
\cite{90 Berkin and Maeda}.
Besides, the gravitational field equation for even higher-order gravity was derived
by imposing the Noether symmetry 
\cite{00 Capozziello and Lambiase}. 
 %
%

In cosmology, modified gravity models are usually applied to the inflationary epoch or to the present age 
in order to describe the accelerated expansion of the universe.
The effects of $f(R)$ gravity was considered also in the radiation-dominated era (RDE) 
for the study of baryogenesis
\cite{06 lambiase and scarpetta}
or big bang nucleosynthesis (BBN) 
\cite{09 kang and pano, 15 kusakabe etal, 16 kusakabe etal, 06 De Felice etal}. 
In particular, besides the numerical solutions for the given gravity models, a standard solution for a scale factor describing an evolution of RDE Friedmann-Lema\^itre-Robertson-Walker (FLRW) universe was found for a theory with a generic Lagrangian containing almost arbitrary function of $R$ 
\cite{83 Barrow and Ottewill}. 
In this paper,
we show that the standard RDE solution is still viable
for the gravity models involving the fourth or sixth order differential equations.
See the $B$ and $C$ terms in Eq.\,(\ref{trace of GFE})
which were proposed in the previous literature.
This proves that, even within modified gravitational theory involving higher-order terms, 
the standard cosmic evolution 
is valid
in the modified gravity containing the higher-order terms.

In Section II, 
we introduce the generalized Einstein field equation for our model.
In Section III, starting with some standard cosmological assumptions,
we apply the gravity models
from specific one with a basic power-law ansatz of the scale factor
to general case 
and find a common standard RDE solution in Eq.\,(\ref{sol}),
which is
our main result.
Section IV is dedicated to two observational aspects of a specific case of the modified theory.
The final section is devoted for the mathematical conclusion and discussions.

\section{Gravitational field equations}
The gravitational field equation (GFE) \cite{90 Gottloeber etal, 97 N and H, 90 Berkin and Maeda} from the metric variation \cite{72Weinberg, 83Barth} of the action (\ref{action}) is
%
%
\bea
&&
\Big[   g_{ab}  \Big(   {\partial f \over \partial R} \Big)^{;c}_{\phantom{;}c}   -  \Big(   {\partial f \over \partial R} \Big)_{;ab}
          + {\partial f \over \partial R}   R_{ab} - {1 \over 2} f g_{ab}  \Big] 
 - 8\pi ( T^{(B)}_{ab}   + T^{(C)}_{ab} ) 
 = 8\pi  T_{ab}^{(m)}   
          \label{GFE}   	,
 \eea 
 where \bea	
 &&   T^{(B)}_{ab}
\equiv
 {B \over {8\pi}} \Big( {1 \over 2}  R^{cd} R_{cd} g_{ab} + R_{;ab} - 2 R^{cd} R_{acbd} - {1 \over 2} g_{ab}  { R^{;c} }_{c} - {{R_{ab}}^{;c}}_{c}  \Big)  {}
                \label{EM tensor B}        ,
\\  && T^{(C)}_{ab} \equiv   
{C \over {8\pi}} \Big[ 2 { R^{;c} }_{c;ab} - 2R_{ab} { R^{;c} }_{c} + R_{,a} R_{,b} 
                               - g_{ab} \big( 2 { {  { R^{;c} }_{c} }^{;d} }_d + {1 \over 2 } R^{;c} R_{,c}  \Big)    \Big]         .
 \eea
For a derivation of the GFE, we 
use the following relation
\bea
&& \delta (\Box R) = 
 - \delta g_{ab} {R^{ab;c}}_{c} - \delta g_{ab} R^{;ab} - g^{ab} \delta  {{g_{ab;c}}^{cd}}_{d}
  +  \delta {{g_{ab}}^{;bac}}_{c}    
 \nonumber \\&& \phantom{\delta (\Box R) =}  
-  R^{ab} \delta {g_{ab;c}}^{c}   - 2 {R^{ab}}_{;c} \delta {g_{ab}}^{;c}
 - R^{;a} \delta {g_{ab}}^{;b}     + {1 \over 2} R^{;c} g^{ab} \delta g_{ab;c}      ,
\eea
referring to the helpful work by Barth and Christensen \cite{83Barth}.
 %
The contraction of the GFE in Eq.\,(\ref{GFE}) with an inverse metric $g^{ab}$ will be useful for RDE cosmology, which is given as
\bea
\Big[  3  { \Big( {\partial f \over \partial R}  \Big)^{;c}}_{c} + {\partial f \over \partial R} R - 2 f   \Big] 
  + 2  B { R^{;c} }_{c}
 + C \big[   6  {{{ R^{;c} }_{c}}^{;d}}_d   + 2 R  { R^{;c} }_{c} + R^{;c} R_{,c}    \big]
= 8 \pi T^{(m)c}_{\phantom{(m)}c} .  \label{trace of GFE}
\eea
For later convenience, we also introduce an alternative equivalent form of the GFE after arranging the directly varied GFE in Eq.\,(\ref{GFE})
 \cite{91H, 97 N and H}
\bea
&& R_{ab} - {1 \over 2} R g_{ab} =   {1 \over F}   
\Big[ 8\pi T_{ab}^{(m)} + 
 F_{;ab} - g_{ab} \Big( {F^{;c}}_{c} - {{f - RF} \over 2}  \Big)    
+ C \Big(  R_{,a} R_{,b} - {1 \over 2} g_{ab} R^{;c} R_{,c}  \Big)
\nonumber \\ && 
\phantom{G_{ab} =}
  + B \Big( {2 \over 3} R R_{ab} +  {1 \over 2}  R^{cd} R_{cd} g_{ab} 
+{ 1 \over 3} R_{;ab}
 - 2 R^{cd} R_{acbd} 
+ {1 \over 6} g_{ab}  { R^{;c} }_{c} 
- {{R_{ab}}^{;c}}_{c}   \Big)
 \Big]      ,
\label{GFE H}
\\ && F \equiv f_{,R} + {2 \over 3} B R + 2 C {R^{;c}}_{c}  ,   
\quad  f_{,R} \equiv   {\partial f \over \partial R} ,
\eea
and the trace of Eq.\,(\ref{GFE H})  
\bea
-R 
= {1 \over F} \big[ 8\pi T^{(m)c}_{\phantom{(m)}c}    
      -3 {F^{;c}}_{c}  + 2 f - 2  F R  
      +{2 \over 3}  B R^2   
       - C R^{;c} R_{,c}  \big]   .
\label{trace of GFE H}    
\eea  
\section{background universe during the RDE}
%
 We adopt a spatially flat FLRW metric 
representing a homogeneous and isotropic background universe
\bea 
 && ds^2 =  a^2(\eta)  \big( -d \eta^2 + \delta_{\alpha \beta} dx^{\alpha} dx^{\beta}   \label{metric} \big) ,
\eea
where $a(\eta)$ is the cosmic scale factor, $\eta$ is the conformal time ($dt \equiv a d\eta$), and $\dot a / a \equiv {1 \over a} { {da} \over {dt} } \equiv H $ is the Hubble expansion rate. 
We 
adopt a perfect fluid in standard cosmology, whose
energy-momentum tensor is
composed of time-dependent energy density $\mu(t)$ and pressure $p(t)$
\bea
T^{(m)0}_{\phantom{(m)}0} = - \mu(t) , 
\quad T^{(m)0}_{\phantom{(m)}\alpha} = 0,
 \quad T^{(m)\alpha}_{\phantom{(m)}\beta} = p(t) \delta^\alpha_\beta   
\label{EM tensor}  .
\eea
In the 
RDE
the cosmological constant is negligible 
and 
the equation of state (EoS) for radiation-like
is simply given 
with the RDE density 
\bea 
p = {1 \over 3} \mu 
\label{EoS} .
\eea
\subsection{Case $N=2, C=0$}
In this subsection, a special case with $ f(R) = R + A R^2 $ (Starobinsky model of $f(R)$ gravity) and $C=0$ (no sixth order gravity) is considered. 
Then the temporal and spatial components of the GFE in Eq. (\ref{GFE}), respectively, become \cite{97 N and H}
\bea
 && 8\pi  \mu = 3 H^2 + 6 (3A+B) \big( 2 \ddot H H - \dot H ^2  +  6 \dot H H^2 \big)  
\label{Friedmann eq} ,
\\ && 8\pi p  = -  ( 2\dot H + 3 H^2) 
  - 2 (3A+B) \Big( 2 { {d^3  H} \over {dt^3}} + 12 \ddot H H + 9 \dot H ^2 + 18 \dot H H^2 \Big)   
\label{accel eq} ,
\eea
where it is notable that $A$- and $B$-terms play qualitatively the same role 
for the evolution of the background universe. 
The modified Friedmann equation in Eq.\,(\ref{Friedmann eq}) and  Eq.\,(\ref{accel eq}) can be checked by substituting them into the
following continuity equation:
\bea
\dot \mu + 3 H (\mu + p ) = 0
\label{conservation eq} .
\eea
A system of three independent ordinary differential equations (ODEs) in Eqs.\,(\ref{Friedmann eq}), (\ref{accel eq}) (or (\ref{conservation eq})), and (\ref{EoS}), contains 3 unknown functions, $ a(t), \mu(t),$ 
and $p(t)$. Putting Eqs.\,(\ref{Friedmann eq}) and (\ref{EoS}) into Eq.\,(\ref{conservation eq}),  
the set of 
ordinary differential equations
can 
be reduced as an ODE for $H(t)$ that has various solutions:
%
\bea 
 \dot H + 2 H^2 
  + 2 (3A+B) \Big( {{d^3  H} \over {dt^3}}  + 7 \ddot{H} H + 4 \dot{ H}^2 +12 \dot H H^2    \Big)  = 0 .        \label{3rd order H}
  \eea
This is a nonlinear third order ODE for $H(t)$ (fourth order eq. for $a(t)$); however, an exact analytical solution can be easily found by a power-law ansatz:
\bea 
a(t) \propto t^{\alpha},
\quad H(t) = {\alpha \over t} ,
\quad \dot H = - { \alpha \over {t^2} } ,  
\quad \ddot H = 2 {\alpha \over {t^3}},
\quad { {d^3  H} \over {dt^3}} = -6 {\alpha \over t{^4}} .
 \label{power-law}
\eea
Among two mathematical solutions, $\alpha = 1/2$ or $\alpha=0$, for Eq.\,(\ref{3rd order H}), the former is selected
 as a standard solution.

By reducing order of the ODE with the definition of the expansion rate (from fourth to third order) and the following chain rule (from third to second order) 
for $X = X(t)$, 
\bea
&& \ddot X = \dot X { {d  \dot{X} }\over {dX} } ,
 \quad  {{d^3 X} \over {dt^3}} = { {dX} \over {dt}  } { d \ddot{X} \over {dX} }  
                                                =  \dot{X} \Big[ \dot X { {d^2 \dot X} \over { dX^2 } }  +  \Big( { {d  \dot{X} }\over {dX} } \Big)^2      \Big]  ,
\\ &&
 Y \equiv \dot X , 
\quad
{ {dY} \over {dX} } =   { {d  \dot{X} }\over {dX} }    = {{ \dot Y} \over {Y}} ,  
\quad
{ {d^2 Y}  \over {dX^2}  }  = {1 \over Y^2 } \Big( \ddot{Y} - { {\dot Y}^2 \over {Y}} \Big)   \label{yprime} ,
\eea
we try to alleviate an instability that the fourth order ODE for $a(t)$ in Eq.\,(\ref{3rd order H}) mathematically suffers. 
Thus, Eq.\,(\ref{3rd order H}) has this mathematical form with no explicit time dependence
\bea
&& Y + 2 X^2 + 2 (3A+B) Y \Big[  
{ {d^2 Y} \over { dX^2 } } Y + \Big( { {dY} \over {dX} } \Big)^2  + 7  { {dY} \over {dX} } X + 4 Y + 12  X^2  \Big]   =   0   
\label{math 2nd order} ,
\eea 
\bea
&& 
 \dot H + 2 H^2  
  + 2 (3A+B) \dot H \Big [ { d^2 \dot{H} \over dH^2} \dot H + \Big({{d \dot H} \over {dH} }\Big)^2 + 7 {{d \dot H} \over {dH} } H + 4\dot H + 12 H^2  \Big]  
 = 0. 
\label{physics 2nd order}
\eea
As one simplest form of solution,
$ \dot H(H) = -2 H^2$ is an exact analytic solution of the second order ODE in Eq.\,(\ref{physics 2nd order}) regardless of the value of $(3A+B)$,
so it can be used to test computer code implementing the corresponding differential equations by comparing analytical plot and the numerical results
\cite{in preparation}.

%

Consequently, the flat FLRW universe model under 
$R + AR^2$ as well as $B R^{ab}R_{ab}$ gravity has a standard RDE solution
\bea
a(t) = (const) t^{1/2},  \quad    H(t) = { 1 \over {2 t }} , 
\quad   \dot H(H) = -2 H^2   \label{RDE solution} ,  \label{sol}
\eea
where the third one describes a flipped parabola (a 2D phase diagram) on a $H$-$\dot H$ plane in which only quadrant IV plane is physically 
acceptable. 
We can also notice that the RDE solution (\ref{sol}) is still
viable in more generalized models such as any polynomial $f(R)$ or a sixth order gravity.
%
\subsection{Case $B=0=C$, any natural number $N$}
For a polynomial $f(R)$ (Eq.\,(\ref{polynomial})) model  setting $B=C=0$,
we use the same method with the previous case to get an ODE for $H(t)$ whose counterpart is Eq.\,(\ref{3rd order H})
\bea
  && \Big[  { {\dot R} \over 2} - 3 \big( \ddot H +  2  \dot H H   \big) \Big] f_{,R}
    - 3 \big( \dot H + H^2  \big) \dot R { {d } \over {dR}} f_{,R}   
    + 3 \dot R { d \over {dR}} \Big(  H \dot R {d  \over {dR}} f_{,R}  \Big) 
\nonumber   \\ && 
 + 4 H \Big[  {  f \over 2 }  - 3 \big(   \dot H + H^2 \big) f_{,R}  + 3 H \dot R { {d} \over {dR} } f_{,R}  \Big]
 = 0 ,   \label{ODE H f(R)} 
\eea
where the Ricci scalar $R$ from the metric in Eq.\,(\ref{metric}) and its time derivative are, respectively
\bea
 && R = 6 \big( \dot H + 2 H^2  \big)  , 
\quad \dot R = 6 ( \ddot H + 4 \dot H H  ) .   \label{Ricci scalar}
\eea
Substituting Eq.\,(\ref{power-law}) into Eq.\,(\ref{Ricci scalar}), one can obtain
\bea 
R(t) = { {6 \alpha (2 \alpha -1)}  \over t^2 } .  \label{Ricci scalar using power-law}
\eea
This result simply shows how the standard RDE solution in Eq.\,(\ref{sol}) from the Einstein gravity ($N=1$) becomes also a solution of the GFE in Eq.\,(\ref{ODE H f(R)}) 
involving
any polynomial $f(R)$ gravity 
because the condition of $\alpha=1/2$ yields
\bea
 R = 0 = \dot R = \ddot R = f, \quad f_{,R} =1 . \eea
According to 
Barrow and Ottewill \cite{83 Barrow and Ottewill},
the standard RDE evolution given in Eq.\,(\ref{sol}) is a solution of not only a polynomial $f(R)$, but also any $f(R)$ theory in which $f(0) = 0$ and $f_{,R}(0) \neq 0$. 
Those exceptional examples are $f(R) \sim R^{-N}, \ln R,$ 
and so on. 

Meanwhile the trace in Eq.\,(\ref{trace of GFE}) in this case $(B=C=0)$ is useful \cite{10 Sotiriou and Faraoni},
 especially for the perfect fluid governed by the RDE EoS in Eq.\,(\ref{EoS}),
\bea 
\Big[  3  { \Big( {\partial f \over \partial R}  \Big)^{;c}}_{c} + {\partial f \over \partial R} R - 2 f   \Big] 
  = T^{(m)c}_{\phantom{(m)}c} = 0 .  \label{trace of GFE f(R)}
\eea 
%
%
%
%
This differential equation equivalent to Eq.\,(\ref{ODE H f(R)}) within the flat FLRW model tells us \cite{10 Sotiriou and Faraoni} 
that 
$f(R)$ theories have more various solutions (the exact solution (\ref{sol}) is just one of them) than Einstein's theory ($f=R$)
relating $R$ with $T^{(m)c}_{\phantom{(m)}c}$ 
not differentially but algebraically,
and that
the function (\ref{sol}) is not necessary but sufficient to be a solution of the ODE in Eq.\,(\ref{trace of GFE f(R)}).

\subsection{General case}
\begin{figure}
\includegraphics[width=10cm]{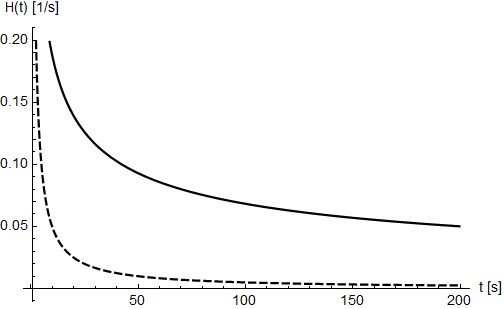}
\caption{
Two solution curves for Eq.\,(\ref{3rd order H}) (Case $N=2, C=0$). 
The analytic solution $H=1/(2t)$ (dashed line) is plotted regardless of $(3A+B)$.
An approximate numerical solution (solid line) is introduced for $(3A+B) = 0.1s^2$. 
}
\label{fig1}
\end{figure}

In analogous to 
the previous cases, it is now easy to see the trace in Eq.\,(\ref{trace of GFE H}) (or Eq.\,(\ref{trace of GFE})) for a RDE flat FLRW universe
\bea
-R 
= {1 \over F} \big[     
      -3 {F^{;c}}_{c}  + 2 f - 2  F R  
      +{2 \over 3}  B R^2   
       - C R^{;c} R_{,c}  \big]   ,
\label{trace of GFE H RDE}    
\eea  
has an analytical solution 
of
$H(t) = 1/(2t)$
because
the solution ($\alpha = 1/2$) implies that
\bea
R = 0 = {R^{;c}}_{;c} = - (\ddot R + 3H \dot R) = f = \dot R, 
\quad  F \equiv f_{,R} + {2 \over 3} B R + 2 C {R^{;c}}_{c} = 1 = f_{,R} .
\label{vanishing Box R}
\eea
The traced Eq.\,(\ref{trace of GFE H RDE}) is a
complicated
fifth order ODE for $H(t)$ admitting various solutions, 
but the standard RDE solution is a sufficient condition to 
satisfy Eq.\,(\ref{trace of GFE H RDE}) regardless of the constants $A_{(n \geq 2)}, B$, and $C$.
Among other various numerical solutions of Eq.\,(\ref{trace of GFE H RDE}) depending on those constants,
it would be an intriguing problem which one could be a candidate as a physical solution describing the evolution of a RDE universe
affected by the modified gravity models.
Figure \ref{fig1} shows one of numerical solutions of Eq.\,(\ref{3rd order H}) for $(3A+B) = 0.1s^2$ 
in the Case $N=2, C=0$ 
.
%
\section{Observational tests for the theory $N=2, C=0$} 
Although we focus on a theory of the homogeneous and isotropic universe model, we would like to briefly mention two observational aspects for the Case $N=2, C=0$ (subsection III. A.), i.e. Starobinsky plus $B R^{ab} R_{ab}$ model upto the linearly perturbed order. Those are BBN in the FLRW background universe and gravitational wave from the disturbed universe.  

Observations of primordial abundances such as $^2{\rm H}$, $^3{\rm He}$, $^4{\rm He}$, and $^7{\rm Li}$ are currently the only data providing information of RDE. Since the change in cosmic expansion rate significantly affects the freeze-out time of the nucleosynthesis, the BBN calculation with the given gravity model could valuate this model \cite{in preparation}. Specifically, for the BBN calculation, the cosmic expansion rate can be obtained by solving Eqs.\,(\ref{Friedmann eq}) and (\ref{accel eq}) with thermodynamic variables of dominant species such as photons, neutrinos, electrons, and positrons. Then, taking the modified cosmic expansion rate, one can carry out BBN calculation with traditionally reliable \cite{kawano code} or user-friendly improved \cite{primat code} BBN codes, whose result would depend on the coefficient $(3A+B)$. Therefore, by the observations of primordial abundances, the coefficient $(3A+B)$ whose original dimension is length to the second power could be constrained
\cite{in preparation}.
In particular, we note that there could be the potential to distinguish between Einstein's theory of gravity 
and generalized gravity theories using precisely measured amounts of helium ($^4{\rm He}$) and deuterium ($^2{\rm H}$).


The propagation speed of tensor-type first order cosmological perturbation can distinguish $B R^{ab} R_{ab}$ from $AR^2$ gravity in principle. If those modified gravity effects are small enough to be treated perturbatively comparing to Einstein gravity, then the wave equation can be simplified to Mukhanov-Sasaki equation that implies the propagation speed slightly less than the speed of light due to only $B R^{ab} R_{ab}$ effect \cite{Y 20}. 
If the speed of cosmological gravitational wave is determined to $c$ in future observations with very high accuracy, then $B R^{ab} R_{ab}$ model will be inappropriate as a gravity theory and $AR^2$ model will be able to survive. While $A$- and $B$- gravity behave similarly at the background order as we see above, they are very distinguishable in the next order both mathematically \cite{97 N and H} and physically. In reality, it is very difficult to detect or measure the subtle interval of arrival between electromagnetic and gravitational wave signal.   
It is currently unclear which gravity model is the better explanation for the RDE universe;
however, we wish to emphasize that our RDE solution for the evolution of the background universe is very available in observationally selected model(s) among the aforementioned Cases.  

%
%
\section{CONCLUSION AND DISCUSSIONS}
We conclude that 
the standard RDE solution of $a(t) \propto t^{1/2}$ obtained from the Einstein gravity $(N=1)$ is also a RDE solution in a spatially flat FLRW universe filled with perfect fluid under a generalized gravity model  
whose Lagrangian is
$ \Big[ {1 \over 16\pi } \big(\sum_{n=1}^{N} A_{(n)} R^n  + B R^{ab} R_{ab} + C R  {R^{;c}}_{c}  \big)  + L_m \Big]  
 $.
Besides the polynomial $f(R)$ theory ($A_{(n)}$-terms) as a subset of more general $f(R)$ gravity \cite{83 Barrow and Ottewill}, 
the fourth order $B R^{ab} R_{ab}$ ($B$ gravity) and sixth order $C R  {R^{;c}}_{c}$ ($C$ gravity) theories in the FLRW model also 
allow the same solution in Eq.\,(\ref{sol}) by virtue of the 
qualitative sameness
of the $R^{ab} R_{ab}$ with the $R^2$ theory
(see Eqs. (\ref{Friedmann eq}, \ref{accel eq}, \ref{3rd order H})) 
 and of the vanishing property of $\Box R$ (Eq. (\ref{vanishing Box R})) with the solution.
This analytical solution can be some fiducial curves (regardless of the coefficients $A_{(n \geq 2)}, B,$ and $C$) that can test the numerical computations to find other numerial solution curves (depending on the coefficients).

At first we made efforts to obtain some numerical RDE solutions of Eq.\,(\ref{3rd order H}) instead of the exact analytical solution in the special case ($R^{ab} R_{ab}$ model added to Starobinsky theory for $N=2$), 
we inductively obtained a hypothesis that the solution in Eq.\,(\ref{sol}) is also a solution in any polynomial $f(R)$ model
and deductively tried to prove it,
to which the discovery for general $f(R)$ gravity by Barrow and Ottewill was prior \cite{83 Barrow and Ottewill}.

%
For the Case $N=2, C=0$ (subsection III. A.), we have either standard analytical solution in Eq.\,(\ref{sol}) or other numerical solutions for the third order ODE in Eq.\,(\ref{3rd order H}). 
If Eq.\,(\ref{sol}) is a real physical description of the evolution of the universe during the RDE, then $A$ and $B$ terms introduced in the action serve theoretically useful purposes (such as quantum corrections or renormalization \cite{77 Stelle}) without affecting the background evolution although the constants $A$ and $B$ are arbitrary with this solution that cannot be determined from observational constraints.
However if other non-standard solutions of scale factor are allowed by the RDE universe governed by the higher-derivative gravity, then the slight effects of $A$ and $B$ terms can be constrained by the comparison of the BBN observations and calculations \cite{in preparation}. 

Even if the terms involved by $A$ and $B$ 
representing the four-derivative theory in the action Eq.\,(\ref{action})
share the same solution at the background order in the homogeneous and isotropic model,
the two terms behave quite differently in the perturbed universe with gravitational wave \cite{97 N and H, Y 20} or density fluctuation \cite{99 N and H}. 
A 
 model with an action containing a general function of a contracted quantity $R^{ab} R_{ab}$
\cite{05 carroll etal} 
is not likely to have the same RDE solution in Eq.\,(\ref{sol})
since  the quantity $R^{ab} R_{ab} = 12 \big( {\dot H}^2 + 3 \dot H H^2 + 3 H^4  \big) $
is non-vanishing even if $H = 1/(2t)$. 
While the generic $f(R)$ and $ R  {R^{;c}}_{c}$ gravity can be conformally transformed into Einstein gravity with single minimally coupled scalar field (MSF) 
\cite{84 Whitt, 10 Sotiriou and Faraoni, 05Mukhanov, 92Mukhanov}
and with two interacting MSFs
\cite{90 Gottloeber etal}
 respectively, 
$R^{ab} R_{ab}$ gravity does not have such a symmetry \cite{97 N and H, 99 N and H}.

How about the standard 
matter-dominated era (MDE) solution $a(t) \propto t^{2/3}$ ? 
This MDE solution from the Einstein gravity is hardly an exact solution in a polynomial $f(R)$ gravity as well as in the $B$ or $C$ gravity 
unless the effects of the $A_{(n \geq 2)}$, $B$ and $C$ theories are small enough during the MDE
because $R(t)$ cannot become zero with $\alpha = 2/3$. 
However, if the universe evolved like Eq.\,(\ref{power-law}) during the RDE, the correction terms from the $B$ or $C$ gravity might decay enough to converge to the standard curve $a(t) \propto t^{2/3}$ (See Appendix).
The MDE in $f(R)$ gravity was investigated in Ref.\,\cite{06 capozziello etal}.
It is also noticed that semiclassical analysis of the Wheeler-DeWitt equation describing the universe before the inflationary epoch allows a radiation-like solution $a(t) \propto t^{1/2}$ \cite{08 Appignani and Casadio}
and that power-law expanding $a \propto t^\alpha$ can be an attractor solution in some fourth-order $f(R)$ models that have conformal symmetry
\cite{89 Schmidt}.

One of our main assumptions is a \emph{spatially flat} FLRW universe (three space curvature $K=0$),
beyond which our argument is unlikely established 
since the Ricci scalar 
\bea 
R = 6 \big( \dot H + 2 H^2  +  { K \over {a^2} } \big)
\eea
 from the FLRW metric including the closed or open universe model
\cite{20 N and H and Barrow, 20 Pitrou etal, 08 P, 17 O'Raifeartaigh etal}
\bea
 ds^2 =  a^2 \Big(  - d \eta^2 + 
{ {dr^2} \over  (1-Kr^2)  } + r^2 ( d\theta^2 + \sin^2 \theta d\phi^2  )   
  \Big) ,  
\eea
is non-vanishing with $a(t) \propto t^{1/2}$
unless $ K / a^2$ term is negligible.
For simplicity, our arguments are based on the homogeneous and isotropic assumption, so called the cosmological principle.
However, there is a debate on the principle
\cite{P etal 17, 22 Kim etal, 05 Hogg etal, 22 Aluri etal}
and researchers 
\cite{amendola and tsujikawa 10, ellis etal 12}
introduce alternative metrics beyond FLRW,
e.g. to explain the current accelerating expansion of the universe.
Comparison between $A$ and $B$ models using those metrics is
another issue.

Cosmological perturbations for the $f(R)$ and $ R  {R^{;c}}_{c}$ gravity
($B=0=L_m$ and general $f(R)$ in Eq.\,(\ref{action}))
were investigated \cite{91H}, 
where the arranged GFE in Eq.\,(\ref{GFE H}) is more convenient than the directly 
varied GFE in Eq.\,(\ref{GFE}) 
since each component of the Einstein tensor $G^a_b \equiv R^a_b - {1 \over 2} R \delta^a_b $ up to the linearly perturbed order is
already calculated and listed, e.g. in Ref.\,\cite{11H}.
The equation for a tensor-type (tracefree and transverse)
 perturbation variable $C^{(t)}_{\alpha\beta}$ in Fourier space
 for the model
was derived as \cite{91H}
\bea 
\ddot C^{(t)}_{\alpha\beta} + \Big( 3H + {\dot F \over F} \Big) \dot C^{(t)}_{\alpha\beta}
+ { {k^2 + 2K} \over a^2 } C^{(t)}_{\alpha\beta} = 0 , \label{GW eq}
\eea
where $ F = f_{,R} + 2 C {R^{;c}}_{c} $,  
and $k$ is the wavenumber of the perturbations.
Interestingly,
this complicated gravitational wave (GW) equation can be simplified 
to a standard form
when the evolution of a flat background universe is described by the 
solution in Eq.\,(\ref{sol})  
and $f_{,R}= 1$ during the relating era
so that $F$ becomes unity. 
Further, 
the simplified GW equation can 
become 
the Bessel equation whose exact solutions are known
\cite{11H, 91H, 05 Peter et Uzan, book08Weinberg}.
The propagation speed of cosmological gravitational wave, in principle, can play a role of a discriminator in selection of gravity models. Most of models predict that the speed is the same as $c$ while a perturbative $B R^{ab} R_{ab}$ model \cite{Y 20} as a subset of more general theory by Weinberg \cite{PRD08Weinberg} and a model \cite{05 H and N} inspired by string theory do not. For more detail, see TABLE II in Ref.\,\cite{05 H and N} for a flat universe in each generalized gravity.  

Future investigations with a part of this higher-order Lagrangian
would be e.g. 
finding other kinds of physical RDE evolution of the universe beyond the standard solution found here,
and trying applications to physics of neutron star(s), to alternative cosmological metrics or to quantum cosmology  
where modified gravity effects may be significant.

\begin{acknowledgments}
This research was supported by a grant from the National Research Foundation of Korea (Grant No. NRF-2020R1A2C3006177, 
Grant No. NRF-2021R1A6A1A03043957, and
2018R1D1A1B07051126).
The work of D.J. was supported by the Institute for Basic Science under IBS-R012-D1.
\end{acknowledgments}

\appendix*
\section{Decaying higher-derivative terms 
during the radiation-matter coexistence and matter-dominated era}
%
Since the universe evolves from the RDE to the MDE, we elaborate the epoch of coexistence of radiation and matter as well as the MDE with two approximate but intuitive arguments for convergence to the standard MDE solution $a \propto t^{2/3}$. 
According to the scalar-type density fluctuation theory, 
the MDE solution in generalized gravity is required to match the observed large scale structure if viscosity of cosmic fluid is negligible \cite{72Weinberg, book08Weinberg}. 
As a research on the BBN in the Brans-Dicke gravity pointed out \cite{06 De Felice etal}, we basically assume in this Appendix that the scale factor should evolve nearly as $t^{1/2}$ during the RDE so that the correction terms due to $A, B$ and $C$ gravity are small enough to keep upto the first order.

Firstly, if the power-law ansatz in Eq.\,(\ref{power-law}) with a slight deviation from the EoS in Eq.\,(\ref{EoS}), $p \simeq \mu/3 $, i.e. $\alpha \simeq 1/2$, is inserted into Eq.\,(\ref{trace of GFE}) or into Eqs.\,(\ref{Friedmann eq}, \ref{accel eq}), then $C$- and $(3A+B)$-term rapidly decay proportionally to $1/t^6$ and to $1/t^4$ respectively
during the RDE.  
 
Secondly, let us narrow down to $A$ and $B$ gravity and treat the higher-derivative terms in Eqs.\,(\ref{Friedmann eq}, \ref{accel eq}) 
by using the standard evolution of the double-component (radiation and matter) flat universe 
described by the dimensionless Friedmann equation
\bea 
{H^2 \over H_0^2} 
= {\Omega_{r0} \over y^4} + {\Omega_{m0} \over y^3}
\quad \Longleftrightarrow \quad
\dot{y}^2 = ( a_0 y )^{-2} [ \alpha_r^2 + 2\alpha_m y ]
 ,   \label{dimensionless Friedmann eq}
\eea
whose integration by part results in $t=t(a)$ form \cite{03 Ryden}:
\bea
H_0 t = 
{4 \over 3}  {y^2_{rm} \over \sqrt{\Omega_{r0}}}  \Big[ 1 - \Big( 1- {1 \over 2} {y \over y_{rm}}  \Big)  \sqrt{1+{y \over y_{rm}} }  \Big] ,
\label{t(a) in r-m era}
\eea
where
$ H_0 \equiv H(t_0) $ is the Hubble constant, $t_0$ is the present time, $y \equiv a(t) / a_{0}$ is the normalized scale factor, 
$\alpha_m \equiv a_0^2  H_0^2 \Omega_{m0} /2 $ 
and 
$ \alpha_r \equiv \big( a_0^2 H_0^2 \Omega_{r0} \big)^{1/2} $
\cite{ellis etal 12}
are constants determinded by observational values,
$\Omega_{i0} \equiv \mu_{i0} / \mu_{c0}$ 
represents the energy density contribution of $i$-component  (e.g. subscripts $r$ and $m$ stand for radiation and matter),
$\mu_{c0} \equiv {3 \over 8\pi} H_0^2 $ is the present critical density,
and $y_{rm} \equiv a(t_{rm}) / a_0 \equiv \Omega_{r0} / \Omega_{m0}$ 
is the normalized scale factor when $\mu_r = \mu_m$ at $t=t_{rm}$. 
Eq.\,(\ref{t(a) in r-m era}) is insightful because its RDE ($a \ll a_{rm}$) and MDE limits ($a \gg a_{rm}$, but until the contribution of the cosmological constant is negligible) recover the results from the standard cosmology 
\cite{03 Ryden}:
\bea
&& a \ll a_{rm} \quad \Rightarrow \quad
 y \simeq \big( 2 \sqrt{\Omega_{r0}}  H_0 t \big)^{1/2},
\\ &&
a \gg a_{rm} \quad \Rightarrow \quad
 y \simeq \Big(  {3 \over 2} \sqrt{\Omega_{m0}} H_0 t  \Big)^{2/3} .
\eea
%
Eq.\,(\ref{t(a) in r-m era}) is also useful because it enables us to
calculate the time of radiation-matter equality when $a=a_{rm}$ \cite{03 Ryden}:
\bea t_{rm} = {4 \over 3} \Big( 1 - {1 \over \sqrt{2}}  \Big) H_0^{-1} { \Omega_{r0}^{3/2} \over \Omega_{m0}^2}
,
\eea
whose value is about 23,000 years
if we use the following 
observational values $H_0 = 67.4 \phantom{.} \mathrm{km/s/Mpc} = 100 h \phantom{.} \mathrm{km/s/Mpc}$, 
\phantom{l}
$\Omega_{m0}=0.315$ determined from Planck observation of the cosmic microwave background (CMB) with the $\Lambda$CDM model \cite{Planck2018 cos params}, and $\Omega_{r0} = 2.47 \times 10^{-5} h^{-2}$ determined by the CMB temperature, $T$ = 2.7255 K \cite{22 Lahav and Liddle, 09 Fixsen}.

However, since Eq.\,(\ref{t(a) in r-m era}) for the radiation plus matter universe has a form of inverse function from which $H(t)$ and its $t$ derivatives are hardly calculated, we had better use the exact solution in terms of the dimensionless conformal time $\eta$ ($dt \equiv a d \eta$) \cite{ellis etal 12}
\bea 
a(\eta) = a_0 \Big[{1 \over 2} \alpha_m \eta^2 + \alpha_r \eta  \Big] ,  \label{a(eta) in r-m era}
\qquad
t = {a_0 \over 6} \big[ \alpha_m \eta^3 + 3 \alpha_r \eta^2  \big] ,
\eea
setting $t=\eta=0$ when $a=0$.
Using Eq.\,(\ref{a(eta) in r-m era}) derived from Einstein gravity
allows us to estimate approximately the decaying behavior of 
the higher-derivative terms in Eqs.\,(\ref{Friedmann eq}) and (\ref{accel eq}), respectively, as $\eta$ elapses,
\bea 
&&
\big( 2 \ddot H H - \dot H ^2  +  6 \dot H H^2 \big)  
= 
{1 \over a^4} \big[ 2 \mathcal{H}'' \mathcal{H} - (\mathcal{H}')^2 - 3 \mathcal{H}^4    \big] 
\nonumber \\ &&
\simeq - {\alpha_m \over  2 { a_0^4 \big( {\alpha_m \over 2} \eta^2 + \alpha_r \eta  \big)^7  } } 
           \Big(  8 \alpha_r^2 + 18 \alpha_r \alpha_m \eta  + 9 \alpha_m^2 \eta^2  \Big)       ,
\label{decaying term in Friedmann eq}
\\ && 
 \Big( 2 { {d^3  H} \over {dt^3}} + 12 \ddot H H + 9 \dot H ^2 + 18 \dot H H^2 \Big)    
\nonumber \\ &&
=
{1 \over a^4} \big[ 2 \mathcal{H}'''  - 2\mathcal{H}'' \mathcal{H} + (\mathcal{H}')^2 - 12 \mathcal{H}' \mathcal{H}^2 + 3 \mathcal{H}^4  \big] ,
\nonumber \\ &&
\simeq 
{ \alpha_m \over {2  a_0^4 } \big( {\alpha_m \over 2} \eta^2 + \alpha_r \eta  \big)^7 }
\Big( 32 \alpha_r^2 + 54 \alpha_r \alpha_m \eta + 27 \alpha_m^2 \eta^2    \Big)    ,
\label{decaying term in accel eq}
\eea
where $\mathcal{H} \equiv a'/a \equiv {1 \over a} { da \over {d\eta} } = a H$.
If there are only subtle deviations from the standard cosmological evolutions in the fourth order gravity, in other words, the terms from Einstein's theory such as $8\pi\mu, 3H^2, 8\pi p$ , and $-(2\dot H  +3 H^2 )$ in Eqs.\,(\ref{Friedmann eq}, \ref{accel eq}) are roughly like
$\propto \eta^{-4}$ and $\propto \eta^{-6}  $ in the RDE and the MDE respectively, 
then  the Eqs.\,(\ref{decaying term in Friedmann eq}, \ref{decaying term in accel eq}) show that how much faster the higher-derivative corrections disappear than the standard terms do as time goes by. 
For a while let us imagine a hypothetical universe filled with only radiation literally i.e. the case of $\alpha_m = 0$, which implies that
Eqs.\,(\ref{decaying term in Friedmann eq}) and (\ref{decaying term in accel eq}) vanish.
This is consistent with the fact that the exact solution in Eq.\,(\ref{sol}) makes those terms nought.
Values of the dimensionless parameter $w(t) \equiv p/ \mu = (p_r + p_m)/ (\mu_r + \mu_m) = 1/[3(1+ y/y_{rm})] $ are nearly 1/3 for the RDE ($y/y_{rm} \ll 1$), 
1/6 for the time of radiation-matter equality ($y/y_{rm} = 1$), and 0 for the MDE ($y/y_{rm} \gg 1$, but before the dark energy effects meaningfully appear), respectively.

In short, the higher-derivative terms rapidly decay to recover the standard cosmological equations that contains the MDE solution 
$a \propto t^{2/3}$ under the condition that the scale factor behavior in the fourth-order gravity subtly deviates from the solution from GR.

If we more narrow down to $A$ gravity (“Starobinsky model”), the theory becomes a subset of $f(R)$ cosmology. In this case, 
we note that there is
a more general exact solution that covers not only the radiation-dust coexistence era \cite{08 Capozziello and De Felice} but also MDE plus dark energy epoch (almost the whole history of cosmic evolution) with a change of the equation of state for $f(R)$ gravity that corresponds to the recent comic acceleration \cite{08 Capozziello etal}. They also pointed out that their solution successfully coincides with the standard MDE solution for redshift interval from 2 to 4 so that the requirement for the formation of large scale structure is fulfilled \cite{08 Capozziello etal}. The problem which specific functions among generic $f(R)$ are selected by astronomical observations is still ongoing.


\end{document}